\documentclass{pasa}%

\usepackage{graphicx}
\usepackage[dvipsnames]{xcolor}
\usepackage{tablefootnote}
\usepackage{amsmath}

\title[High Cadence Optical Transient Searches]{High Cadence Optical Transient Searches using Drift Scan Imaging III: Development of an Inexpensive Drive Control System and Characterisation and Correction of Drive System Periodic Errors}

\author[]{Steven Tingay
\affil{International Centre for Radio Astronomy Research, Curtin University, Bentley, WA 6102, Australia}%
}%

\jid{PASA}
\doi{10.1017/pas.\the\year.xxx}
\jyear{\the\year}

\usepackage{aas_macros}
\usepackage{hyperref} 
\hypersetup{colorlinks,citecolor=blue,linkcolor=blue,urlcolor=blue}

\begin{document}

\begin{frontmatter}
\maketitle

\begin{abstract}
In order to further develop and implement novel drift scan imaging experiments to undertake wide field, high time resolution surveys for millisecond optical transients, an appropriate telescope drive system is required.  This paper describes the development of a simple and inexpensive hardware and software system to monitor, characterise, and correct the primary category of telescope drive errors, periodic errors due to imperfections in the drive and gear chain.  A model for the periodic errors is generated from direct measurements of the telescope drive shaft rotation, verified by comparison to astronomical measurements of the periodic errors.  The predictive model is generated and applied in real-time in the form of corrections to the drive rate.  A demonstration of the system shows that that inherent periodic errors of peak-to-peak amplitude $\sim$100$''$ are reduced to below the seeing limit of $\sim$3$''$.  This demonstration allowed an estimate of the uncertainties on the transient sensitivity timescales of the prototype survey of \citet{2021PASA...38....1T}, with the nominal timescale sensitivity of 21 ms revised to be in the range of 20 $-$ 22 ms, which does not significantly affect the results of the experiment.  The correction system will be adopted into the final version of high cadence imaging experiment, which is currently under construction.  The correction system is inexpensive ($<$\$A100) and composed of readily available hardware, and is readily adaptable to other applications.  Design details and codes are therefore made publicly available.
\end{abstract}

\end{frontmatter}

\section{Introduction}
\label{sec:introduction}
Previously, \citet{2020PASA...37...15T} and \citet{2021PASA...38....1T} (hereafter Papers I and II, respectively) developed a new technique for high cadence optical imaging aimed at the detection of short timescale transients in wide-field surveys.  The technique utilises the motion of the sky across a sensor attached to a wide-field optical system to achieve high time resolution (21 ms in Papers I and II), with the optical system actively driven in an east-west direction at a higher than sidereal rate.

Paper I described a pre-prototype system and proof of concept observations.  Paper II described survey observations obtained with a prototype system, with a transient duration sensitivity of 21 ms and a V magnitude upper limit of 6.6.  An upper limit on the transient event rate of 0.8 events per square degree per day was estimated from 24 hours of observation.  These results are competitive with the results from other recent experiments \citep{2020PASJ...72....3R, 2021AJ....161..135A}.

Beyond the prototype, the construction of a larger-scale instrument is underway, with commissioning due to commence in the second half of 2021.  This instrument builds upon the prototype with larger aperture and wider field optical systems (two Celestron 279mm aperture Rowe-Ackermann Astrographs), full frame imaging sensors (two Canon RP full frame cameras), and the possibility for access to a wider range of transient timescales by virtue of a more flexible drive system.  The aim is to achieve transient timescale sensitivities much closer to the notional 1 ms timescale for objects such as Fast Radio Bursts (FRBs; \citet{2018NatAs...2..845B}).

The development and characterisation of the drive system is the subject of this paper, continuing to describe the development of the overall system.  As the high cadence imaging technique depends on higher than sidereal rate tracking to achieve high time resolution, knowledge of and control of the drive rates is very important.  Telescope drive systems of various levels of complexity require different levels of pointing corrections.  At the professional end, highly complicated models are required to correctly point and track a large telescope \citep{1974PhDT.......206H}.  At the amateur commercial end, many relatively inexpensive modern telescopes include a Periodic Error Compensation (PEC) system \citep{2012JRASC.106..162S}.

For the work under consideration here, which deals with optical systems of moderate size and is highly focused on the exploitation of inexpensive commercial-off-the-shelf (COTS) equipment, periodic errors in the drive system (motor and gear trains) are the dominant form of error.  Periodic errors are generally caused by imperfections of various types in the manufacture and meshing of the gear trains that transfer the motion of a motor to the rotation of the telescope right ascension axis.  The broad effect is the periodic speed-up and slow-down of the motion of the telescope, relative to the sidereal rate, causing an astronomical object to wander back and forth on a sensor, in an east-west (right ascension) direction.  For the high cadence imaging application, the effect will cause a variable transient sensitivity timescale.  This should be understood and, ideally, corrected in the drive system.

The goal of the work described here is to develop, test, and demonstrate a simple and inexpensive system to characterise and correct periodic errors in the drive systems that will be used for the high cadence imaging application.  A system of hardware and software, based on the popular low cost Arduino platform and python code, is developed that allows periodic errors to be measured and characterised, independently of astronomical characterisation.  The measurements are verified by astronomical observations and are used to create a predictive model for the correction of periodic errors.  

This system is developed using the mount and drive components used in Paper II, which allows an estimate of the uncertainties due to periodic errors that affected the experiment described in that paper.  The system is also developed in order to be adapted to the final high cadence imaging instrument currently under construction.  In addition, this work utilises other hardware components selected for the final system, such as the Canon RP cameras, allowing experience to be gained with those components.

The hardware and software system could be readily adapted to many other situations, so design and construction details are provided here and all codes are available at https://github.com/steven-tingay/High-Cadence-Imaging-III.

In Section \ref{sec:hardware} the hardware components of the system are described.  In Section \ref{sec:obsdataproc}, the measurements and astronomical observations used to test and verify the performance of the system are described.  In Section \ref{res} the results are described and discussed, including an estimate of uncertainties for the experiment described in Paper II.  Finally, in Section \ref{conc}, conclusions and future work is briefly described.
 
\section{Hardware under development and test}
\label{sec:hardware}
A simple system has been chosen, based on very inexpensive off-the-shelf parts available at any hobbyist electronics store and openly available software libraries for standard computing platforms.


The drive motor is driven and monitored via an Arduino-based system, in order to measure periodic errors and feed corrections back to the drive system.  The detected and corrected periodic errors are independently verified via direct astronomical measurements, again using simple hardware.

\subsection{Optical assembly, mount, and camera}
\label{optical}
The classic method of measuring periodic error in astronomical telescope systems is via direct astronomical measurements \citep{2012JRASC.106..162S}.  At the end of the day, this is the measurement that matters, the drift of the position of an astronomical object across the sensor, as a function of time.  Thus, these measurements are the ultimate source of verification for any periodic error correction system.

A simple optical assembly is used, a Skywatcher 130mm aperture, 1000mm focal length reflector.  With a spherical primary and a catadioptric corrector before the secondary, the off-axis imaging quality is poor.  But for a substantial on-axis area, stellar images are point-like and easily adequate for the measurement of periodic errors.

As the Canon RP full-frame sensor camera was selected for the final high cadence imaging system, this camera was utilised here, in order to gain general experience with it.  The Canon RP was mounted to the optical assembly (with $x$ and $y$ sensor axes aligned north-south and east-west, respectively) and produced good results.  The sensor pixel size of 5.75$\mu$m corresponds to 1$''$.18 with the optical assembly (on-axis).  The camera was controlled via a MacBook Pro laptop, running the Canon EOS Utility (version 3)\footnote{hk.canon/en/support/0200593702}.

\subsection{Arduino-based drive control and measurement}
\label{arduino}
The telescope mount drive motor is a Nippon Pulse Motor Company PF42-48f3g stepper motor\footnote{www.dynetics.eu/CMS/Docs/NPM/NPM stepper cat.pdf} with 48 steps/revolution and a 200:1 gearhead, giving 0.0375 deg/step.  The further gear reduction of 180:1 provided by the worm and worm gear assembly of the mount gives 0.$''$75/step driving of the right ascension axis of the mount.  The nominal sidereal tracking rate corresponds to one rotation of the worm every 8 minutes, which in turn corresponds to a rotation rate of the motor of 25 RPM.  

The PF42-48f3g motor is a bipolar stepper with the rotor driven by two coils, with two leadwires per coil, which connect to an Arduino Uno (Rev 3) board \footnote{store.arduino.cc/usa/arduino-uno-rev3} and an Arduino Compatible Motor Servo Controller Module\footnote{www.jaycar.com.au/medias/sys\_master/images/ images/9505700151326/XC4472-manualMain.pdf}.  Control of the motor was then enabled by writing an Arduino sketch based on the AFMotor library\footnote{learn.adafruit.com/afmotor-library-reference/af-stepper-class} and supporting python3 scripts.  All code for this work is available from https://github.com/steven-tingay/High-Cadence-Imaging-III.

In order to have access to accurate timing information, a Real Time Clock (RTC) module\footnote{www.jaycar.com.au/medias/sys\_master/images/ images/9485758693406/XC4450-manualMain.pdf} was connected to the Arduino board.  The RTC was accessed in the sketch via the DS1307RTC libray \footnote{www.arduino.cc/reference/en/libraries/ds1307rtc/} and the Time library\footnote{www.arduino.cc/reference/en/libraries/time/}.

Finally, in order to make direct measurements of the periodic errors in the system, a rotary encoder was attached to the worm shaft and connected to the Arduino board.  The rotary encoder utilised is a quadrature encoded 20 step digital encoder\footnote{www.jaycar.com.au/medias/sys\_master/images/ images/9506700066846/XC3736-manualMain.pdf}.  The quadrature encoding is generally utilised to sense the direction of rotation.  However, at low rotation speeds (such as the 8 minute period of the worm), the quadrature encoding can be conveniently used to detect four sub-steps between each digital step.  Thus, the rotary encoder produces 80 sub-steps per rotation, corresponding to 4.5 degree steps of the worm and, due to the 180:1 gear reduction of the worm and worm gear assembly, 90$''$ increments in right ascension.  No special libraries are required to access the rotary encoder data, as the quadrature encoded measurements are simply read on two of the Uno digital pins using standard methods.

The sketch is loaded to the Uno from a MacBook Pro laptop.  The RTC is synchronised to laptop time at each execution of the sketch.  Due to memory and processing limitations on the Uno board itself, data collected by the Uno are processed both on the Uno and the laptop.  Data and processing results are passed back and forth between the Uno and the laptop via a python3 script, utilising the pyserial library\footnote{pypi.org/project/pyserial/}.

\begin{figure}[!ht]
\begin{center}
\includegraphics[width=0.45\textwidth]{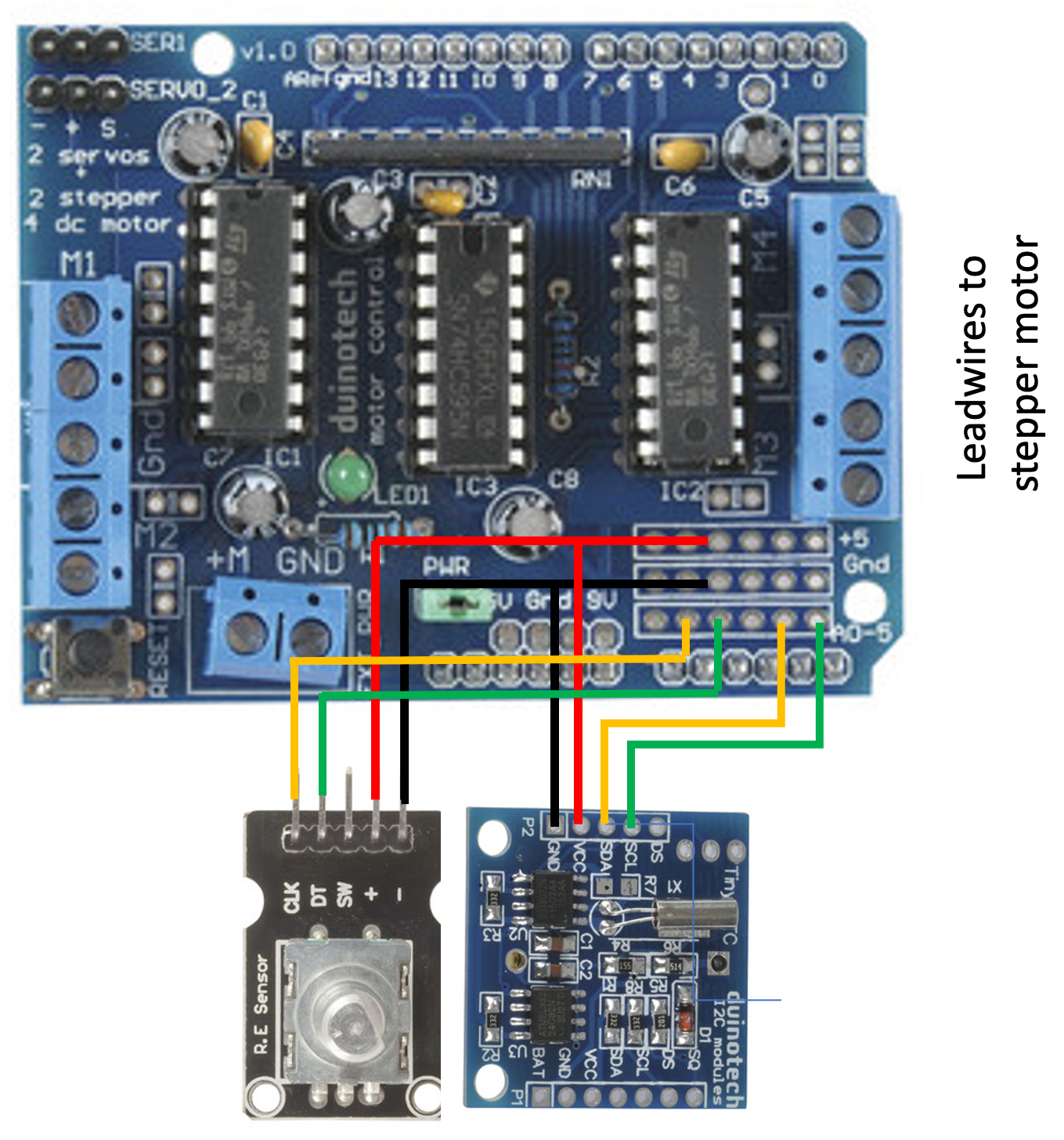}
\caption{The schematic wiring diagram for the Arduino-based drive monitor/controller, showing the motor servo controler board (which mates to the top of an Arduino Uno board), the rotary encoder (bottom left), and the real time clock (bottom right).  The port that connects to the stepper motor is indicated, but wiring is not shown, as this will be dependent on the motor selected.  The quadrature encoded signals from the rotary encoder are carried on pins CLK and DT.  If these signals are unreliably read, the addition of a 100 nF capacitor between each these pins and GND will generally alleviate this issue. A 9V power connection to the Uno and the USB connection to the laptop on the Uno are not shown.}
\label{schematic}
\end{center}
\end{figure}

A schematic wiring diagram for the motor, Uno, Servo Controller Module, RTC, and rotary encoder is provided in Figure \ref{schematic}.  All of the Arduino components cost, in total, less than \$A100.

\section{Measurements and Data Processing}
\label{sec:obsdataproc}

\subsection{Astronomical measurements}
\label{astro}
The most direct measurement of periodic error in a telescope system comes from astronomical measurements.  Thus, as a mechanism to understand the magnitude of the errors that the correction system needs to address, these measurements were undertaken first.

Utilising the mount and camera system described above, and running the Arduino-based motor control system for nominal sidereal tracking, a bright star was acquired on-axis and at the centre of the camera sensor.

Over two rotations of the worm, corresponding to 16 minutes, images of one second exposure length were obtained every five seconds, resulting in 192 images over the 16 minutes.

The bright star was the brightest object in the field-of-view, thus a simple python script (pe.py; included in the github repository for the project) was developed to detect the star in each image and record its sensor location in pixel coordinates (converted to relative arcseconds), along with the time of measurement.  

The script uses the rawpy\footnote{https://pypi.org/project/rawpy/} python module to read the Canon CR3 RAW format images, demosiac them to produce images in the R, G, and B channels corresponding to the Bayer filter (interpolating over the missing pixel values for each colour), and combine the normalised R, G, and B images into a grayscale image. Thus, every pixel in the grayscale image is the result of one measured value and two interpolated values, which means the physical pixel resolution (5.75$\mu$m $\sim$ 1.$''$2) can be used in the analysis.

From the greyscale images, the coordinates of the pixel with the peak value in each image is recorded and utilised as the estimated position of the star as a function of time.

\begin{figure}[!ht]
\begin{center}
\includegraphics[width=0.45\textwidth]{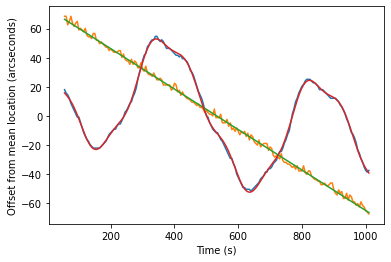}
\includegraphics[width=0.45\textwidth]{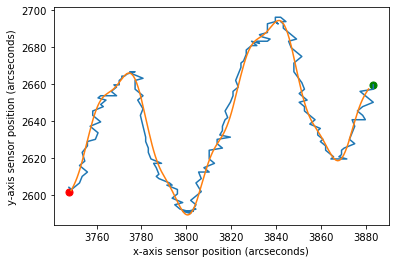}
\caption{Top panel: The drift of the star on the sensor in the north-south (x-axis) direction (orange points and green linear fit) and in the east-west (y-axis) direction (blue points and red sinusoidal fit), as described in the text.  Bottom panel: Motion of the star on the sensor in the ($x$,$y$) domain.  The motion starts at the green marker and the motion ends at the red marker.}
\label{opt}
\end{center}
\end{figure}

A typical example of the results obtained from these measurements can be seen in Figure \ref{opt}, where the top panel shows the drift of the star in the north-south and east-west directions and the bottom panel shows the ($x$,$y$) sensor domain drift as functions of time.

The north-south drift is seen as linear with time, resulting from a very small misalignment of the mount with true north.  This is not a concern and is easily addressed with a small correction to the mount.  However, the drift is advantageous as it makes it easy to inspect what is going on in the east-west direction, which is of primary interest as the direction affected by periodic errors.

In the east-west direction, a small linear drift can be seen over time, corresponding to a small error in the nominal sidereal rate of the drive, superimposed with a large amplitude periodic error, with peak-to-peak variation of approximately 80 $-$ 100 arcseconds and period close to the 8 minute period of the worm.  A random jitter in position due to seeing, of approximately 3 $-$ 5 arcseconds, is also apparent.  This is largely what is expected from a drive system of this quality and the observing location (same as for Papers I and II).

While of large amplitude, the periodic errors are smooth and repeatable over a worm rotation.  Inspection also hints at multiple periodic error components.  In order to characterise both north-south and east-west motions, guided by the visual inspection, python code using the scipy curvefit module\footnote{https://docs.scipy.org/doc/scipy/reference/generated/ scipy.optimize.curve\_fit.html} was used to fit functions to these motions, as follows:

\begin{equation}
D=r_{D}t+o_{D}
\end{equation}
\begin{multline}
R=r_{R}t+o_{R}+a\sin{(bt+c)}+ \\ +d\sin{(et+f)}+g\sin{(ht+i)}
\end{multline}

where: $D$ is the position on the sensor in the declination direction (x-axis), in arcseconds; $r_{D}$ is the linear drift rate in the declination direction, in arcseconds per second; $o_{D}$ is the offset in the declination direction in arcseconds.  This describes a linear drift in the north-south direction.  Also, $R$ is the position on the sensor in the right ascension direction (y-axis), in arcseconds; $r_{R}$ is the linear drift rate in the right ascension direction in arcseconds per second; $o_{R}$ is the offset in the right ascension direction in arcseconds; $a$, $d$, and $g$ are the amplitudes of three sinusoidal periodic errors, in arcseconds; $b$, $e$, and $h$ are the corresponding angular frequencies of the the three sinusoids i radians per seconds, $\frac{2\pi}{T}$, where $T$ is the sinusoid period in seconds; and $c$, $f$, and $i$ are the corresponding sinusoid phase offsets, in radians.  This describes a linear drift with three superimposed sinusoids in the east-west direction.

Figure \ref{opt} shows that these functions for the north-south and east-west motions describe the data very well.  As expected, the highest amplitude sinusoid has an angular frequency corresponding to the nominal 8 minute worm rotation period.  The other two sinusoid periods appear at twice and four times the primary angular frequency.  Table \ref{optfit} lists the fit parameters corresponding to the data in Figure \ref{opt}.  Over many tests, the angular frequencies were very consistent and were thus fixed in all subsequent fitting of data, at 0.0131 rad/s, 0.0262 rad/s, and 0.0524 rad/s, respectively.

\begin{table*}[h!]
\caption{Parameters for fit to north-south and east-west motion astronomical measurements}
\centering
\begin{tabular}{|c|c|c|c|c|c|c|} \hline \hline
$r_{D}$    &$o_{D}$&$r_{R}$&$o_{R}$&$a$/$d$/$g$&$b$/$e$/$h$&$c$/$f$/$i$ \\

$''$/s&$''$            &     $''$/s&$''$                  &$''$             &rad/s&rad           \\ \hline

$-$0.139&3889&$-$0.059     &2673                     &41.5/$-$4.2/4.6                &0.013/0.052/0.026                  &2.82/$-$5.80/$-$7.06                  \\ \hline
\end{tabular} \\
\label{optfit}
\end{table*}

\subsection{Drive shaft encoder measurements}
\label{drive}
The rotary encoder attached to the worm shaft is intended to directly measure any variations in the rotation of the worm that drives the worm gear attached to the right ascension axis.  The intent of these measurements is to find an empirical relationship between the worm shaft phase and the motion of the star on the sensor, in an east-west direction.  If a relationship is found, the rotary encoder data can be used to predict corrections to the drive motor, without further reference to astronomical data.

As the rotary encoder only measures the phase of the shaft as a function of time (with an arbitrary phase offset), the measurement captures the combined effects of all elements in the gear chain, including the gearhead in the stepper motor, the worm, and worm gear (in so far as it affects the worm).  An analysis to isolate the individual periodic errors in each transmission element is well beyond the scope of this work and this is not attempted.  Only the empirical aggregate effect of the periodic errors is considered.

\begin{figure}[!ht]
\begin{center}
\includegraphics[width=0.45\textwidth]{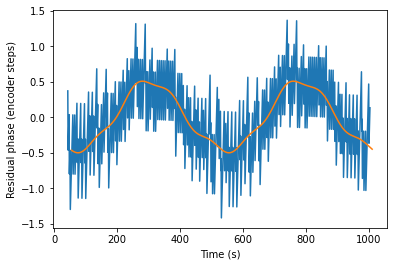}
\caption{The phase of the worm shaft, as measured with the rotary encoder, with vertical axis in units of the encoder steps, over the same period as the astronomical data shown in Figure \ref{optfit}.  Data are shown in blue, the fitted model from Equation 3 is shown in orange.}
\label{rotenc}
\end{center}
\end{figure}

A typical example of the rotary encoder data is shown in Figure \ref{rotenc}, corresponding to the same observation period as the astronomical data shown in Figure \ref{opt}, after the sidereal tracking rate has been removed from the data.  Similar to the astronomical data, a periodic error with angular frequency corresponding to the 8 minute worm period is evident.  While the data appear noisier than the astronomical data, due to the fact that the encoder allows only 80 steps per worm period, some evidence of multiple periods are present in the data.

An identical approach to fitting the astronomical data was adopted for the rotary encoder data, as follows:

\begin{multline}
P=r_{P}t+o_{P}+j\sin{(kt+l)}+ \\ +m\sin{(nt+p)}+q\sin{(st+u)}
\end{multline}

where the parameters are described by analogy to those in Equation 2.  $P$ is the phase in steps of the rotary encoder; $r_{P}$ is the linear rate of the rotary encoder in steps per second; $o_{P}$ is an arbitrary offset in steps (upon every restart of the sketch, the step phase is set to zero); $j$, $m$, and $q$ are the sinusoid amplitudes, in steps; $k$, $n$, and $s$ are the corresponding sinusoid angular frequencies in radians per second, fixed at values of $k=0.0131$, $n=0.0524$, and $s=0.0262$, as noted above; and $l$, $p$, and $u$ are the corresponding sinusoid phase offsets in radians.  Table \ref{rotenc} gives the fitted parameters, similar to those given for the astronomical data in Table \ref{optfit}, and the function is shown in Figure \ref{rotenc}.

\begin{table*}[h!]
\caption{Parameters for fit to rotary encoder measurements}
\centering
\begin{tabular}{|c|c|c|c|c|} \hline \hline
$r_{P}$&$o_{P}$&$j$/$m$/$q$&$k$/$n$/$s$&$l$/$p$/$u$ \\

steps/s&steps                  &steps             &rad/s&rad           \\ \hline

$-$0.166     &6.783                     &$-$0.483/0.036/0.097                &0.0131/0.0524/0.0262                  &0.716/0.225/0.654                  \\ \hline
\end{tabular} \\
\label{optfit}
\end{table*}

Care was taken to ensure that the sense of rotation measured by the rotary encoder matched the sense of the motions on the sensor measured astronomically, so that the two datasets could be compared and examined to determine the relationship between them.

\section{Results and discussion}
\label{res}
In order to compare the data collected from the rotary encoder and the astronomical measurements, the rotary encoder units of steps need to be converted into the astronomical measurement units of arcseconds.  This is expressed as:

\begin{equation}
P_{\rm arcsec}=\frac{1.296\times10^{6}P_{\rm steps}}{S G}
\end{equation}

where: $P_{{\rm(arcsec}}$ is the displacement on the sky in arcseconds caused by $P_{\rm{steps}}$ of the rotary encoder; $1.296\times10^{6}$ is the number of arcseconds in 360 degrees; $S$ is the number of steps per rotation of the rotary encoder (80 in this case); and $G$ is the gear ratio of the worm and worm gear assembly (180:1 in this case).

\begin{figure}[!ht]
\begin{center}
\includegraphics[width=0.45\textwidth]{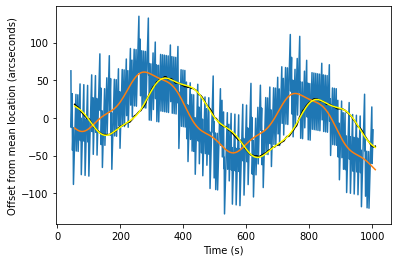}
\caption{The astronomical measurements of the periodic error, shown in black, with the fitted model of Equation 2 shown in yellow.  The rotary encoder data, shown in blue, with the fitted model according to Equation 3 in orange, after conversion to astronomical units via Equation 4, is also shown.}
\label{comp}
\end{center}
\end{figure}

If this conversion is applied to the data in Figure \ref{rotenc} and compared to the data in Figure \ref{optfit}, Figure \ref{comp} results.  In this case, the data and the fit of the data for the rotary encoder have been de-trended for the sidereal rate and have both had the east-west drift rate derived from the astronomical data applied, to provide ease of visual comparison.

\begin{figure}[!ht]
\begin{center}
\includegraphics[width=0.45\textwidth]{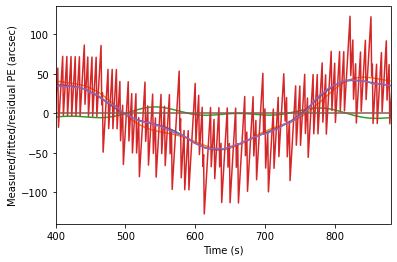}
\caption{The rotary encoder measurements (red), fitted model for the rotary encoder data (red line), model for the astronomical measurements (blue line), and the difference between the rotary encoder model and the astronomical model (green line), as a function of time for an 8 minute period corresponding to one worm rotation.  The rotary encoder measurements and model have been shifted by the 80 second offset from the astronomical model.}
\label{comp2}
\end{center}
\end{figure}

A good match of the rotary encoder data and the astronomical data is evident, with a phase offset between the two datasets.  To examine the offset, both datasets were de-trended of any linear drift in the east-west direction and only the periodic errors were considered.  Using the {\it correlate} function in the python numpy module, the offset was determined to be 80 seconds.  With this offset removed, Figure \ref{comp2} shows the comparison between both datasets and both functional fits to the data, over a single 8 minute rotation of the worm.

Figure \ref{comp2} confirms an excellent match between a prediction based on the rotary encoder data and the measured astronomical effects, with an 80 second timing offset.  Over many tests and trials, the 80 second offset is robustly repeatable (with variations of $\pm$2 seconds), but the origin of the timing offset is unknown within the overall system of gears.

Figure \ref{comp2} also shows the difference between the prediction from the rotary encoder and the astronomical measurements, which represents an expectation for how well the periodic errors can be corrected.  From an uncorrected peak-to-peak amplitude of approximately 100 arcseconds, the corrected data should be able to achieve a peak-to-peak amplitude of approximately 10 arcseconds.

The realisation of these corrections is dependent on a data processing scheme that performs the rotary encoder measurements, fits the three sinusoids to the rotary encoder data, and then derives corrections from these fits and applies them to the drive motor rate, to compensate for the periodic errors.

This is achieved via the Arduino sketch and a python script that collects and processes the rotary encoder data, derives the corrections, and then controls the drive motor appropriately.  Certain limitations on the levels of processing possible on the board, and limitations of the various libraries utilised had to be overcome.  

For example, the Arduino Uno only has 2048 bytes of on-board storage for code and variables.  Thus, collections of rotary encoder data are passed to the laptop via serial communications, where the functional fit is performed.  The fit parameters are sent back to the Uno board, where they are utilised to derive rate corrections for the drive motor, from the time derivative of Equation 3 along with the known time offset, as follows:

\begin{multline}
\frac{dP_{\rm PE}}{dt}=jk\cos{(k(t-t_{\rm off})+l)}+ \\ +mn\cos{(n(t-t_{\rm off})+p)}+ \\ +qs\cos{(s(t-t_{\rm off})+u)}
\end{multline}

The linear east-west drift term is omitted from the correction ($P_{\rm PE}$ only describes the periodic error component), as it is separately dealt with as a constant factor in the Arduino sketch, which also depends on other aspects of the code implementation, such as the adjustments required to maintain a constant update period for the corrections.  The corrections in Equation 5 only deal with the periodic error component.

An issue with the motor control library is that the motor can only be controlled in integer values of RPM.  The drive rate variation required for this application is generally in the range 23 to 27 RPM (centred on the sidereal rate of 25 RPM) and requires smooth adjustment.  A scheme by which the motor drive rate is updated once every second at an integer rate is adopted, with the residual between the exact required rate and the integer rate added to the exact required rate for the next update second, works well, but adds complexity to the code.  In order to maintain a one second update period, a variable number of stepper motor steps per update is calculated, according to the integer RPM in utilisation for that step.

In this way, the overall sidereal tracking rate is maintained with a regular update rate, modulated by the rate change corrections predicted by the rotary encoder data.  The many details of the codes are best described in the comments included in the codes themselves, available via the github repository at https://github.com/steven-tingay/High-Cadence-Imaging-III.

The periodic error correction scheme, and the code, was tested as follows.  The hardware was setup as described above.  A bright star was acquired and tracked over a 1920 second period, corresponding to four rotations of the worm.  Upon execution of the sketch and the python code, the collection of 960 seconds of rotary encoder data was commenced.  At the same time, the collection of 1920 seconds of images was commenced via the optical assembly and camera.  At the end of the first 960 seconds, the model derived from the rotary encoder data was transmitted to the Arduino board and thereafter the board generated drive rate corrections from the model and delivered them to the motor.

\begin{figure}[!ht]
\begin{center}
\includegraphics[width=0.45\textwidth]{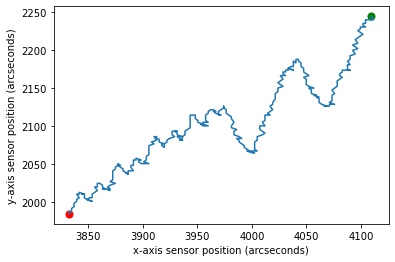}
\includegraphics[width=0.45\textwidth]{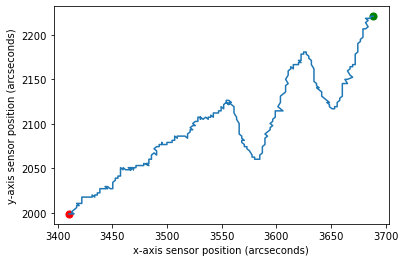}
\caption{The results of verification tests of the periodic error correction process and software.  The top panel shows the results of a 960 calibration period and a further 960 second period when the derived corrections are applied, as seen in astronomical measurements.  The bottom panel shows an identical test at a later point in time, but with only half of the amplitude of the quarter period component of the correction applied, giving a superior result to the top panel.}
\label{comp3}
\end{center}
\end{figure}

Thus, the astronomical measurements cover the first 960 seconds (uncorrected periodic errors) and the second 960 seconds (corrected periodic errors).  The top panel of Figure \ref{comp3} shows the results of this test, in the ($x$,$y$) domain, confirming that the peak-to-peak amplitude in the east-west direction is reduced to the expected approximate 10 arcseconds.

However, following the application of the corrections, even though the periodic errors are greatly reduced, some low level periodicity remains apparent after correction.  The angular frequency of this residual periodic error matches the fitted component at one quarter the period of the worm.  Thus, the test observations were re-run with the application of only half the amplitude of this sinusoid component applied via the corrections.  The result is seen on the bottom panel of Figure \ref{comp3}, where it is apparent that the residual periodic error has been removed.  The RMS around the line of best fit in the period when corrections are applied is 3$''$.3, which is consistent with the seeing from the observing location and larger than the uncertainty due to the pixel size of 1.$''$2.  Why this reduction in the amplitude of this component of the correction is beneficial is not clear, but may be due to the coarse nature of the rotary encoder steps emphasising variability at this angular frequency.

It is possible that further investigation and trial and error may uncover the reasons for this improvement, or allow refinements to this technique.  For example, using more than two worm periods to derive the corrections will likely make the corrections more accurate.  And improving the timing of the system (both camera and rotary encoder times are only good to approximately one second) may help improve the corrections.  Finally, in Figure \ref{comp3} the slow drifts in the north-south and east-west directions have not been rectified, as they represent only very small mechanical adjustments to the mount and very small adjustments to the nominal sidereal tracking rate in software, respectively.  However, the bottom panel of Figure \ref{comp3} shows that the corrections are now very close to seeing limited.  Thus, the objective of the exercise, to obtain an empirical correction of the periodic errors, has been achieved.

\subsection{Tracking uncertainties for published results}
\label{puberr}
One motivation for undertaking this work was to quantify in detail the tracking errors relating to previously published work in Papers I and II.  In particular, Paper II published the first scientific results from the drift scan imaging technique, to place constraints on the event rate of optical transients with V magnitude less than 6.6.  The quoted transient duration the experiment was sensitive to was 21 ms.

This sensitivity timescale was derived from the rate of the sky motion across the pixels of the sensor, when driving the mount at four times the sidereal rate in the anti-sidereal direction.  Clearly, variations in the drive rate can affect this timescale.

As the mount and motor used here was also used in Paper II, these effects can be quantified.  The drive variations experienced in Paper II are exactly those shown in Figure \ref{opt} and tabulated in Table \ref{optfit}.  Using the derivative of Equation 2 (similar to Equation 5) and the coefficients from Table \ref{optfit}, the maximum variation from the sidereal rate can be calculated as 0.$''$9 per second, corresponding to 6\% of the sidereal rate.

Thus, the 21 ms quoted timescales in Paper II can be up to 1 ms in error, when the periodic error gradient is highest.  For the vast majority of the time, the timescale errors are significantly less than 1 ms.  Therefore, the results presented in Paper II are not significantly affected, with the upper limits on the events rates now strictly described as for transients in the range of 20 - 22 ms, rather than for 21 ms.

\section{CONCLUSIONS AND FUTURE WORK}
\label{conc}

A simple and inexpensive method, utilising readily available and inexpensive COTS equipment (total hardware cost of $<$\$A100), has been implemented in order to monitor and characterise periodic errors in telescope drive systems of moderate quality.  Utilising a popular development platform, in the form of Arduino-based hardware and software and python code, the correction of periodic errors to below the seeing limit is demonstrated for the drive system used in Paper II, allowing an estimate of the transient sensitivity timescales for that experiment; rather than the quoted 21 ms timescales, a range of 20 - 22 ms is strictly applicable to that experiment.  The conclusions of Paper II are not significantly affected.

The periodic error correction system will be adapted for the final version of the high cadence imaging experiment, which is currently under construction and due for commissioning in the second half of 2021.

The hardware and software developed for this purpose can be readily adapted for a range of similar applications, so the design details are provided.  All codes utilised for the system are also available in a github repositary at https://github.com/steven-tingay/High-Cadence-Imaging-III.  While all hardware components selected for this project are inexpensive, the methods and codes can be adapted for different and/or higher quality hardware components, for example higher quality rotary encoders with many more than 80 steps.

\begin{acknowledgements}
This research has made use of NASA’s Astrophysics Data System.
\end{acknowledgements}

\bibliographystyle{pasa-mnras}
\bibliography{custom}

\begin{thebibliography}{}
\makeatletter
\relax
\def\mn@urlcharsother{\let\do\@makeother \do\$\do\&\do\#\do\^\do\_\do\%\do\~}
\definecolor{darkblue}{rgb}{0,0,0.597656}
\def\mndoi{\begingroup\mn@urlcharsother \@ifnextchar [ {\mndoi@} {\mndoi@[]}}
\def\mndoi@[#1]#2{\def\@tempa{#1}\ifx\@tempa\@empty \href
  {http://dx.doi.org/#2} {\textcolor{darkblue}{doi:#2}}\else \href
  {http://dx.doi.org/#2} {\textcolor{darkblue}{#1}}\fi \endgroup}
\def\mn@eprint#1#2{\mn@eprint@#1:#2::\@nil}
\def\mn@eprint@arXiv#1{\href {http://arxiv.org/abs/#1} {{\tt arXiv:#1}}}
\def\mn@eprint@dblp#1{\href {http://dblp.uni-trier.de/rec/bibtex/#1.xml}
  {dblp:#1}}
\def\mn@eprint@#1:#2:#3:#4\@nil{\def\@tempa {#1}\def\@tempb {#2}\def\@tempc
  {#3}\ifx \@tempc \@empty \let \@tempc \@tempb \let \@tempb \@tempa \fi \ifx
  \@tempb \@empty \def\@tempb {arXiv}\fi \@ifundefined
  {mn@eprint@\@tempb}{\@tempb:\@tempc}{\expandafter \expandafter \csname
  mn@eprint@\@tempb\endcsname \expandafter{\@tempc}}}

\bibitem[\protect\citeauthoryear{{Arimatsu}, {Tsumura}, {Usui}, {Ootsubo}  \&
  {Watanabe}}{{Arimatsu} et~al.}{2021}]{2021AJ....161..135A}
{Arimatsu} K.,  {Tsumura} K.,  {Usui} F.,  {Ootsubo} T.,   {Watanabe} J.-i.,
  2021, \mndoi [\aj] {10.3847/1538-3881/abd94d}, \href
  {https://ui.adsabs.harvard.edu/abs/2021AJ....161..135A} {161, 135}

\bibitem[\protect\citeauthoryear{{Burke-Spolaor}}{{Burke-Spolaor}}{2018}]{2018NatAs...2..845B}
{Burke-Spolaor} S.,  2018, \mndoi [Nature Astronomy]
  {10.1038/s41550-018-0630-x}, \href
  {https://ui.adsabs.harvard.edu/abs/2018NatAs...2..845B} {2, 845}

\bibitem[\protect\citeauthoryear{{Hovey}}{{Hovey}}{1974}]{1974PhDT.......206H}
{Hovey} G.~R.,  1974, PhD thesis, Australian National University

\bibitem[\protect\citeauthoryear{{Richmond} et~al.,}{{Richmond}
  et~al.}{2020}]{2020PASJ...72....3R}
{Richmond} M.~W.,  et~al., 2020, \mndoi [\pasj] {10.1093/pasj/psz120}, \href
  {https://ui.adsabs.harvard.edu/abs/2020PASJ...72....3R} {72, 3}

\bibitem[\protect\citeauthoryear{{Saunders}}{{Saunders}}{2012}]{2012JRASC.106..162S}
{Saunders} R.,  2012, \jrasc, \href
  {https://ui.adsabs.harvard.edu/abs/2012JRASC.106..162S} {106, 162}

\bibitem[\protect\citeauthoryear{{Tingay}}{{Tingay}}{2020}]{2020PASA...37...15T}
{Tingay} S.,  2020, \mndoi [\pasa] {10.1017/pasa.2020.7}, \href
  {https://ui.adsabs.harvard.edu/abs/2020PASA...37...15T} {37, e015}

\bibitem[\protect\citeauthoryear{{Tingay} \& {Joubert}}{{Tingay} \&
  {Joubert}}{2021}]{2021PASA...38....1T}
{Tingay} S.,  {Joubert} W.,  2021, \mndoi [\pasa] {10.1017/pasa.2020.53}, \href
  {https://ui.adsabs.harvard.edu/abs/2021PASA...38....1T} {38, e001}

\makeatother
\end{thebibliography}

\end{document}